\documentclass[final]{svjour2}

\usepackage{graphicx}  % needed for figures
\usepackage{dcolumn}   % needed for some tables
\usepackage{bm}        % for math

\usepackage{amssymb}
\usepackage[ampersand]{easylist}

\usepackage{color}
\usepackage{setspace}
\usepackage{amsmath}
\usepackage{ulem}

\usepackage{epstopdf}
\usepackage{notes2bib}
\usepackage[caption=false]{subfig}

\begin{document}

\title{Turbulence as information}

\author{W.I. Goldburg   \and
        R.T. Cerbus %etc.
}

\institute{W.I. Goldburg \at
              Department of Physics and Astronomy, University of Pittsburgh, 3941 O'Hara Street, Pittsburgh PA 15260 \\
%              Tel.: +123-45-678910\\
%              Fax: +123-45-678910\\
              \email{goldburg@pitt.edu}           %  \\
           \and
           R.T. Cerbus \at
              Fluid Mechanics Unit, Okinawa Institute of Science and Technology, Onna-son, Okinawa 904-0495, JAPAN
}

\date{}

%\author{W.I. Goldburg}
%\email{goldburg@pitt.edu}
%\affiliation{Department of Physics and Astronomy, University of Pittsburgh, 3941 O'Hara Street, Pittsburgh PA 15260}
%\author{R.T. Cerbus}
%\email{rory.cerbus@oist.jp}
%\affiliation{Fluid Mechanics Unit, Okinawa Institute of Science and Technology, Onna-son, Okinawa 904-0495, JAPAN}

%\begin{affiliations}
% \item Put institutions in this environment and
% \item separate with \verb|\item| commands.
%\end{affiliations}

\maketitle

\begin{abstract}

A message of any sort can be regarded as a source of information. Claude. E. Shannon showed in the last century that information (``what we don't already know'' \cite{gershenfeld2000}) is equivalent to the entropy as defined in statistical mechanics \cite{shannon1948,shannon1964}.  A string of experimental observations is like a succession of words; they both convey information and can be characterized by their entropy. For the fluid flow measurements and simulations to be discussed here (pipe and soap film flow, GOY model), the entropy depends on controllable parameters such as the Reynolds number. The information theory approach is applicable to measurements of any type including those governed by intractable equations or systems where the governing equations are not known. This contribution is dedicated to the memory of Leo Kadanoff, an inspiring teacher and one of the most important scientific leaders of the last half century. 

\end{abstract}

\section{Introduction}

Turbulence is governed by the nonlinear equations of fluid dynamics, but those equations are too complex to be solved analytically.  An understanding of the behavior of turbulent flows is better approached using statistical ideas, while also taking into account the few exact results derived by considering isotropic, homogeneous and incompressible turbulence \cite{landau1959,tritton1988,tennekes1972}. In all cases one must invoke statistical averaging, even if the starting point is the Navier-Stokes equations themselves.  

In this study statistics and probabilities alone appear; the underlying equations of fluid dynamics do not enter at all. Although a probabilistic approach is taken, it is clear that "...'turbulent' does not mean random. The complex motion of the fluid contains characteristic patterns, events and structures that show through all the randomness" \cite{kadanoff1991}. Our aim is to unveil this structure of turbulence using tools from information theory \cite{gershenfeld2000}. Nevertheless one obtains results that are consistent with those deduced using the traditional approach. These include the existence of a cascade, implying the correlation between eddies of differing sizes. 

Let the velocity of the fluid $U$ be a random variable and $u$ be one of its possible values. For simplicity of discussion we consider only a single component. There is an associated probability density function $p(u)$, which is determined experimentally by counting occurrences. Instead of the velocity we could also consider the vorticity measured at a point or a velocity difference measured between two nearby points in a homogeneous fluid. It will be assumed that the turbulent fluid is in a steady state, making the absolute observation time $t$ of little interest.

Once $p(u)$ is in hand, we could proceed to calculate such quantities as the velocity moments $\langle u^n \rangle = \int_{u \in U} p(u) u^n du$. When $u$ is replaced with the velocity difference over a scale $r$: $\delta u(r) \equiv u(x+r) - u(x)$, these moments are called the structure functions \cite{frisch1995}. They play an important role in turbulence theory, with particular prominence given to Kolmgorov's famous exact solution $\langle \delta u(r)^3 \rangle = -\frac{4}{5} \epsilon r$, where $\epsilon$ is the energy injection rate \cite{frisch1995,landau1959}.

Instead of moving in this traditional direction, we proceed in another. We ask what the probability density function itself has to say. Completely random systems have flat distributions, where all possible values have equal weight: $p(u) \sim $ constant. In the other extreme, the distribution is a delta function $p(u) \sim \delta(u - u_0)$, with only one possible value $u_0$. This latter scenario corresponds to the ideal laminar state with no fluctuations. The former case is not seen even when the flow is turbulent, because constraints on the energy of the flow force it to be closer to gaussian. Now we ask whether there is a way to make this description more quantitative.

At the center of information theory is a quantitative measure of the ``broadness" of probability distributions \cite{shannon1948,shannon1964,cover1991}. Shannon called this the entropy, apocryphally encouraged to do so by von Neumann because he could safely hide inside the confusion this term evokes in the physics community \cite{tribus1971}. The most basic form of the entropy, which we shall later revise, is simply 
\begin{equation}
H(U) = - \sum_{u \in U} p(u) \log_2 p(u)
\label{eq:H}
\end{equation}
where $\log_2 p(u)$ can be thought of as a measure of surprise at observing any particular $u$. The form of Eq. \ref{eq:H} is the same as for the energy states in, $e.g.$ the canonical ensemble of statistical mechanics \cite{pathria2011}. The astute reader will notice we have used a sum $\sum$ instead of an integral $\int$. Shannon's original work was on messages with discrete variables (like the letters in the English language), and while there is a generalization to continuous variables \cite{cover1991}, we will switch to the discrete form here. (The discretization of continuous variables needs to be done carefully, and we leave this issue for later.)

Clearly $H(U)$ is large when $p(u)$ is broad and small if it is narrow. Large uncertainty means large $H(U)$. If we are measuring the velocity $u$ in a laboratory and adopt the interpretation that the readout on our instrument is the fluid system's ``message" to us, then $H(U)$ is the amount information we obtain from our measurement. We are told nothing new by repeatedly making measurements of laminar flow, since it never changes, but we are always getting new information from turbulent flow. R. S. Shaw was the first to recognize this information production as a general feature of chaotic systems \cite{shaw1981}.

\begin{figure}[h!]
%\vspace{-1em}
\centering
\includegraphics[scale = 0.3]{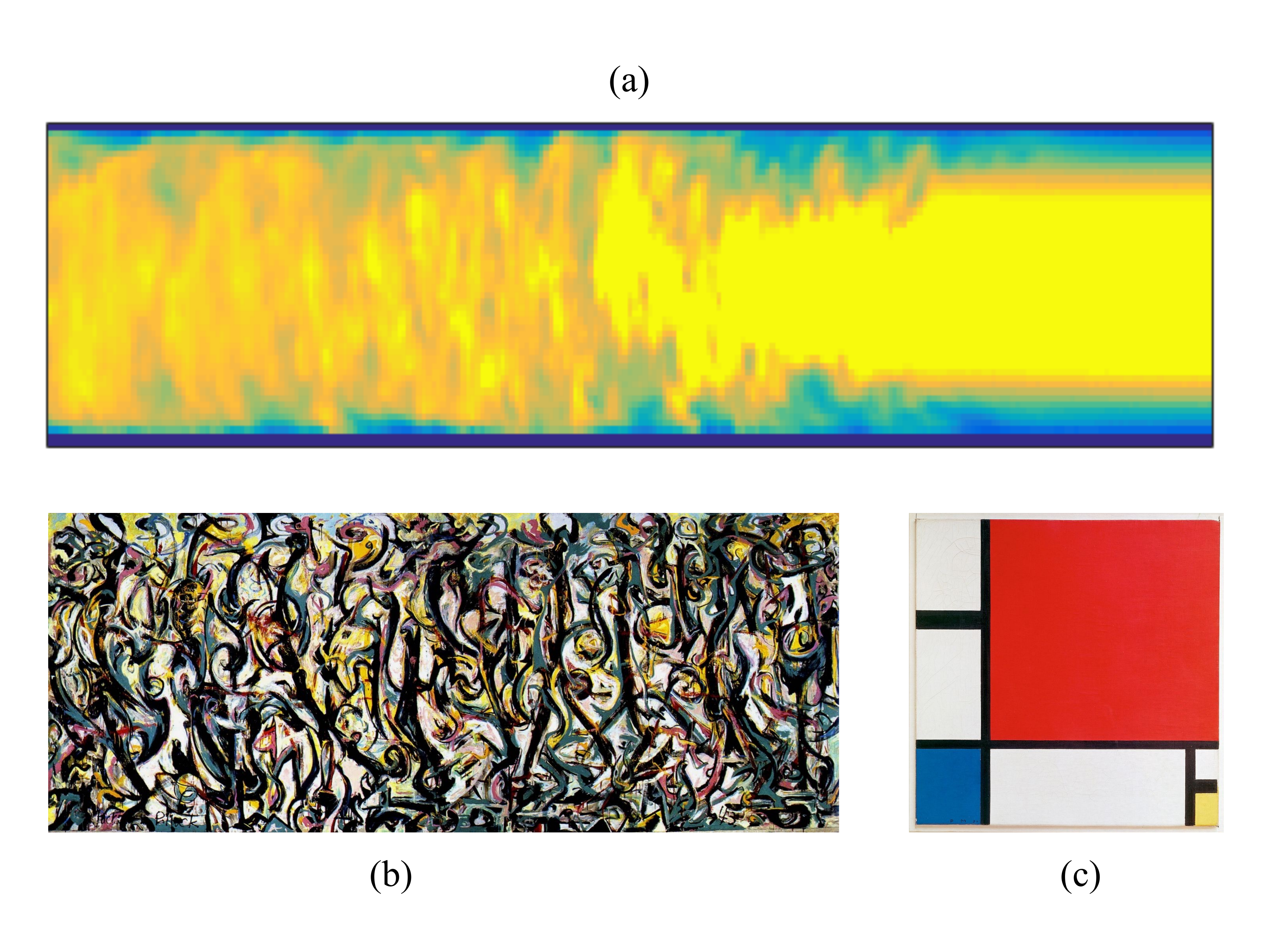}
%\vspace{-3.5em}
\caption{Instantaneous axial velocity field in a pipe with turbulent ($(a)$, left) and laminar flow ($(a)$, right), juxtaposed with Jackson Pollock's {\it Mural} ($(b)$, \cite{pollock}) and Piet Mondrian's {\it Composition With Red, Blue and Yellow} ($(c)$, \cite{mondrian}). Both turbulence and the Pollock painting show a large degree of disorder, while the laminar and Mondrian painting are very ordered. The $Re = 5300$ pipe flow is measured using 2D Particle Imaging Velocimetry. Further experimental details can be found in Sec. \ref{section:pipe}.}
\label{pollock}
\end{figure}

An analogous observation outside of fluid dynamics is an image taken from a newspaper. The image is created by varying the local density of black dots of ink. Just as some images are characterized by wild variations in paint color (a Jackson Pollock painting is a good example), others, such as many Mondrian paintings, show small variations, and have simple geometrical forms. The information or $H$ is larger in the first of these. It is closer to high Reynolds number flows and the Mondrian painting is akin to flow at moderate or low Reynolds number. This visual link is displayed in a juxtaposition of these paintings with the velocity field in a pipe in Fig. \ref{pollock}.

If information theory's utility were limited to the above quantification of uncertainty through $H$, it would indeed be of limited use. However, $H$ is the basis for extracting other quantities, like the mutual (or shared) information $I(X;Y)$ between two quantities $X$ and $Y$ \cite{cover1991}.  These quantities will be introduced in the forthcoming examples as the need arises. The emphasis here is on the application of these tools to real data taken in the laboratory or calculated on a computer.

\section{Transitional flow in a pipe}
\label{section:pipe}

\begin{figure}[h!]
%\vspace{-1em}
\centering
\includegraphics[scale = 0.3]{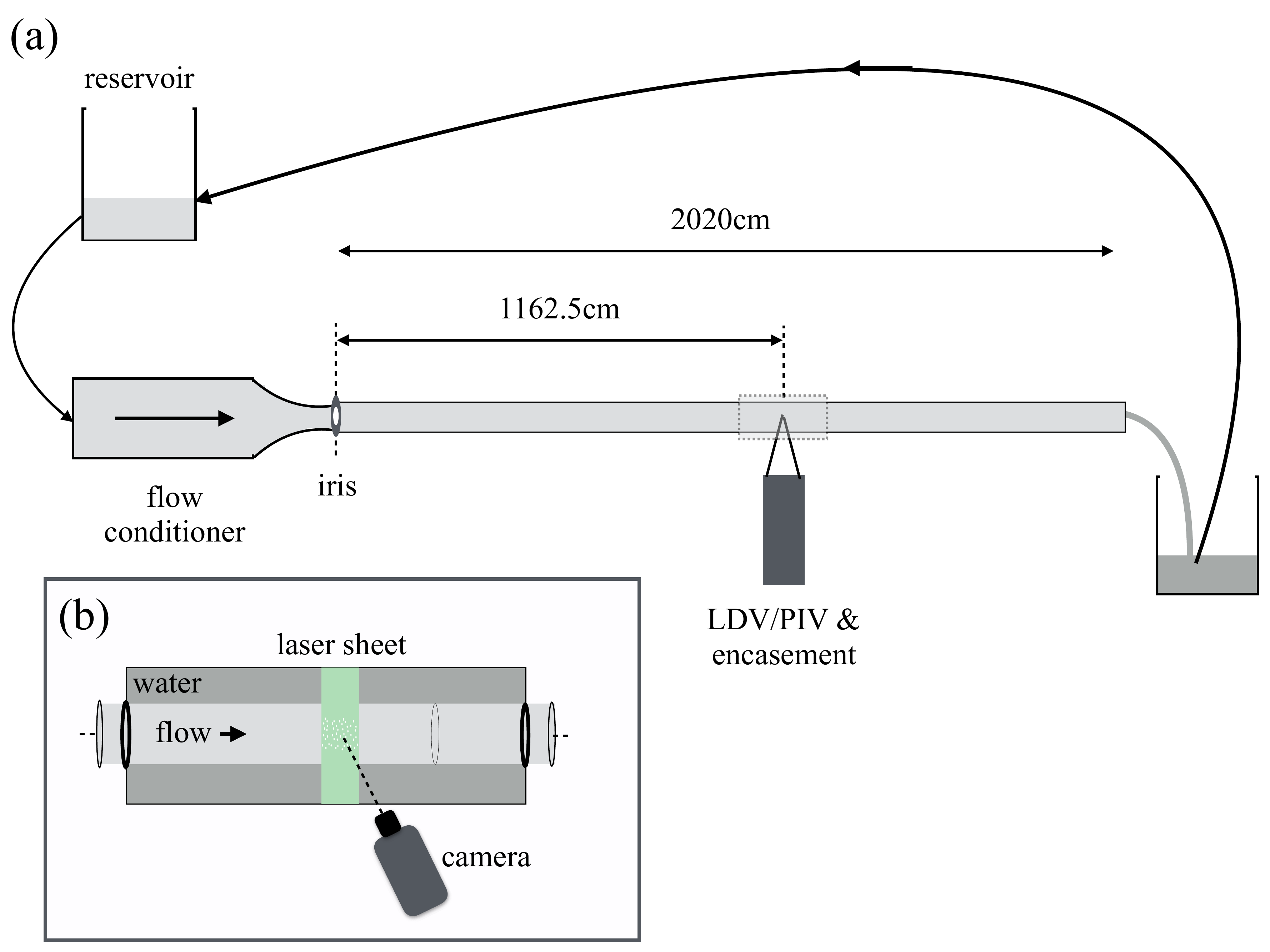}
%\vspace{-3.5em}
\caption{Schematic of the pipe setup (a) and the PIV setup (b). Arrows indicate the direction of the flow, which recirculates. The gravity-driven flow rate is controlled by adjusting the height of a reservoir and with a ball valve. A temperature controller and flow conditioner, along with the smoothness (roughness $< 10\mu$m) of the pipe ensures that the flow remains laminar unless an obstace, an iris \cite{durst2006}, is used to perturb it. The LDV and PIV (b) measurements were taken at a distance of $\approx 1160$ cm from the iris. For the PIV measurements a thin ($< 0.3$ mm) light sheet pierces the center of the pipe from the top and the scattered light is viewed from the side by a Phantom high speed camera. This section of the pipe is encased in a transparent box filled with water to reduce optical distortion.}
\label{pipe_setup}
\end{figure}    

\begin{figure}[h!]
%\hspace{-1.5em}
\centering
\includegraphics[scale = 0.4]{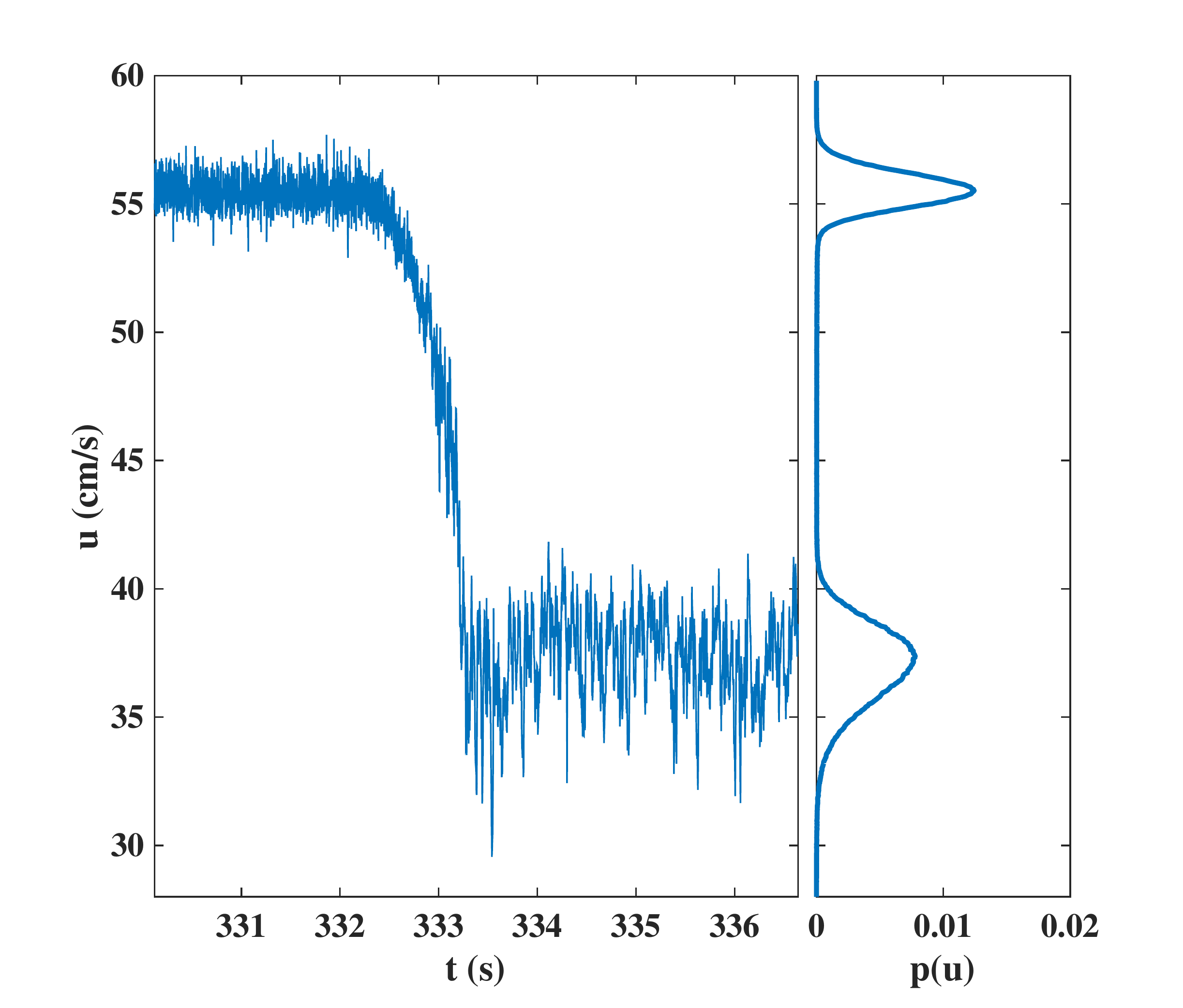}
\caption{An example time series of flow with both laminar and turbulent states on the left, and the corresponding $p(u)$ on the right. The upper value is the laminar and the lower is turbulent. Here $Re = 7000$. Although laminar flow is in principle fluctuation free, instrumental random noise is always present. The experimental settings (LDV SNR, etc.) were chosen to avoid filtering out this noise for the purpose of discussion.}
\label{slug}
\end{figure}

We turn first to demonstrating experimentally the assertion made earlier about the entropy of laminar and turbulent flow. We perform measurements of the axial velocity $u$ at the centerline of a long ($\approx 2000$ cm), cylindrical pipe as a function of time using a Dantec laser Doppler velocimeter (LDV) and hollow glass, silver-coated 10 $\mu$m particles for scattering. The Reynolds number here is defined as $Re = UD/\nu$, where $U$ is the cross-sectionally averaged velocity, and $D = 2.5$ cm is the diameter, with $\nu$ the kinematic viscosity of water. A schematic of the setup is shown in Fig. \ref{pipe_setup}. In pipes, as in many other shear flows, the system can in principle remain laminar to infinite $Re$ \cite{mullin2011}. However, turbulence can be triggered  by a finite perturbation, such as an obstacle. Doing so results in an intermittent time series of turbulent and laminar patches \cite{mullin2011}. Fig. \ref{slug} shows an interval of time when the flow at a point is transitioning between the two states.

The higher value is the laminar state and the lower is the turbulent. The pdfs to the right of the time series confirm that even the laminar flow has some fluctuations (either instrumental or from the setup), although they are smaller than the turbulent ones. While the entropy is theoretically zero for laminar flow, since there is only one value of the velocity, the presence of noise hinders a direct experimental confirmation of this.

We begin by determining $p(u)$. Binning data to make a histogram of inevitably finite experimental or numerical data is a familiar procedure. In Fig. \ref{slug}, very fine bin sizes were used so that to the naked eye the curve looks continuous. There is no problem using these same bin sizes for calculating $H(U)$, but it should not come as a surprise that the bin size choice affects $H(U)$. 

\begin{figure}[h!]
\hspace{-1.5em}
\centering
\includegraphics[scale = 0.4]{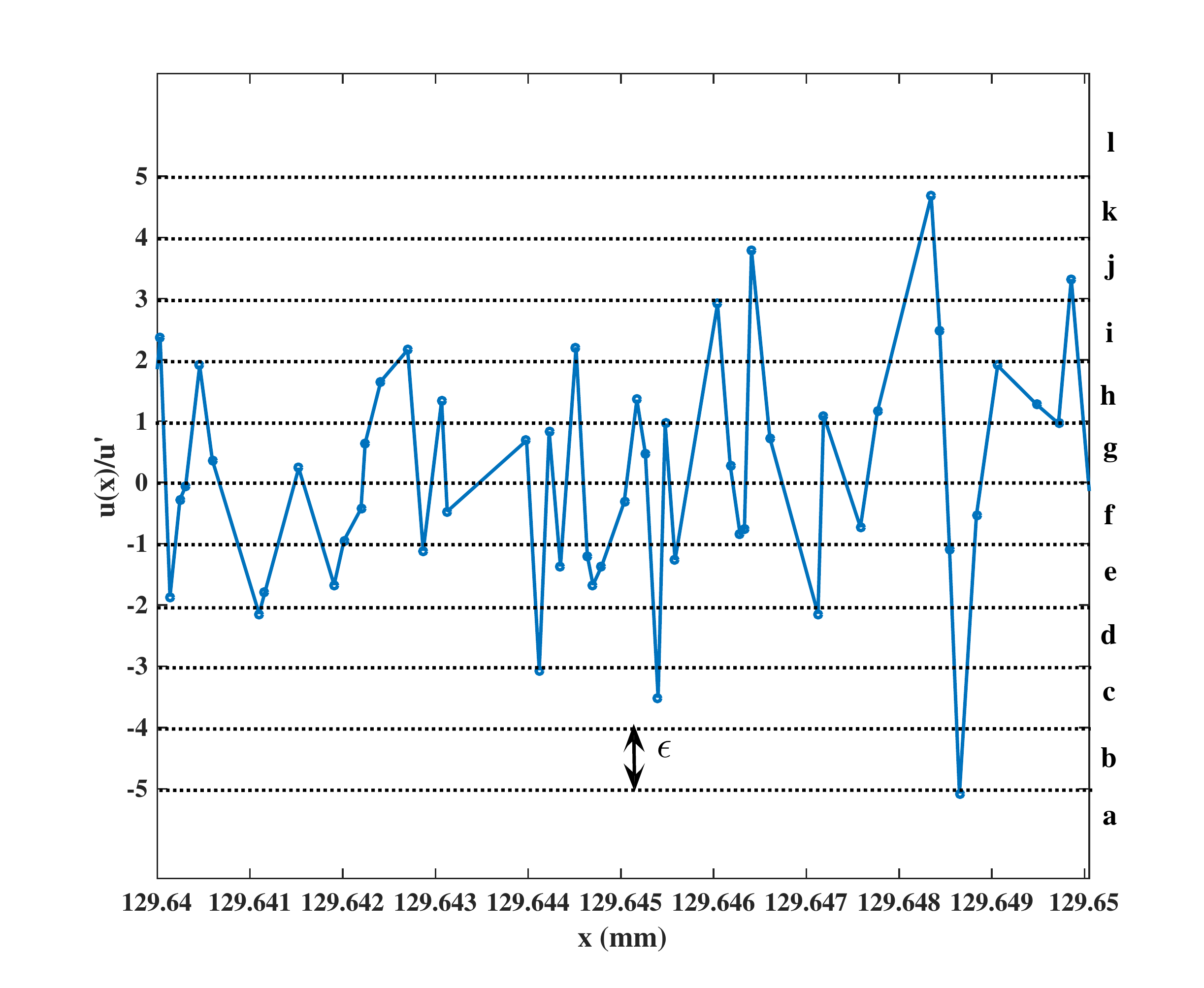}
\caption{An example of a turbulent velocity times series and the boundaries of the bins or segments used to create an alphabet. Any data point that falls in between the dotted lines is assigned the letter of that bin.}
\label{partition}
\end{figure}

\begin{figure}[h!]
\hspace{-1.5em}
\centering
\includegraphics[scale = 0.4]{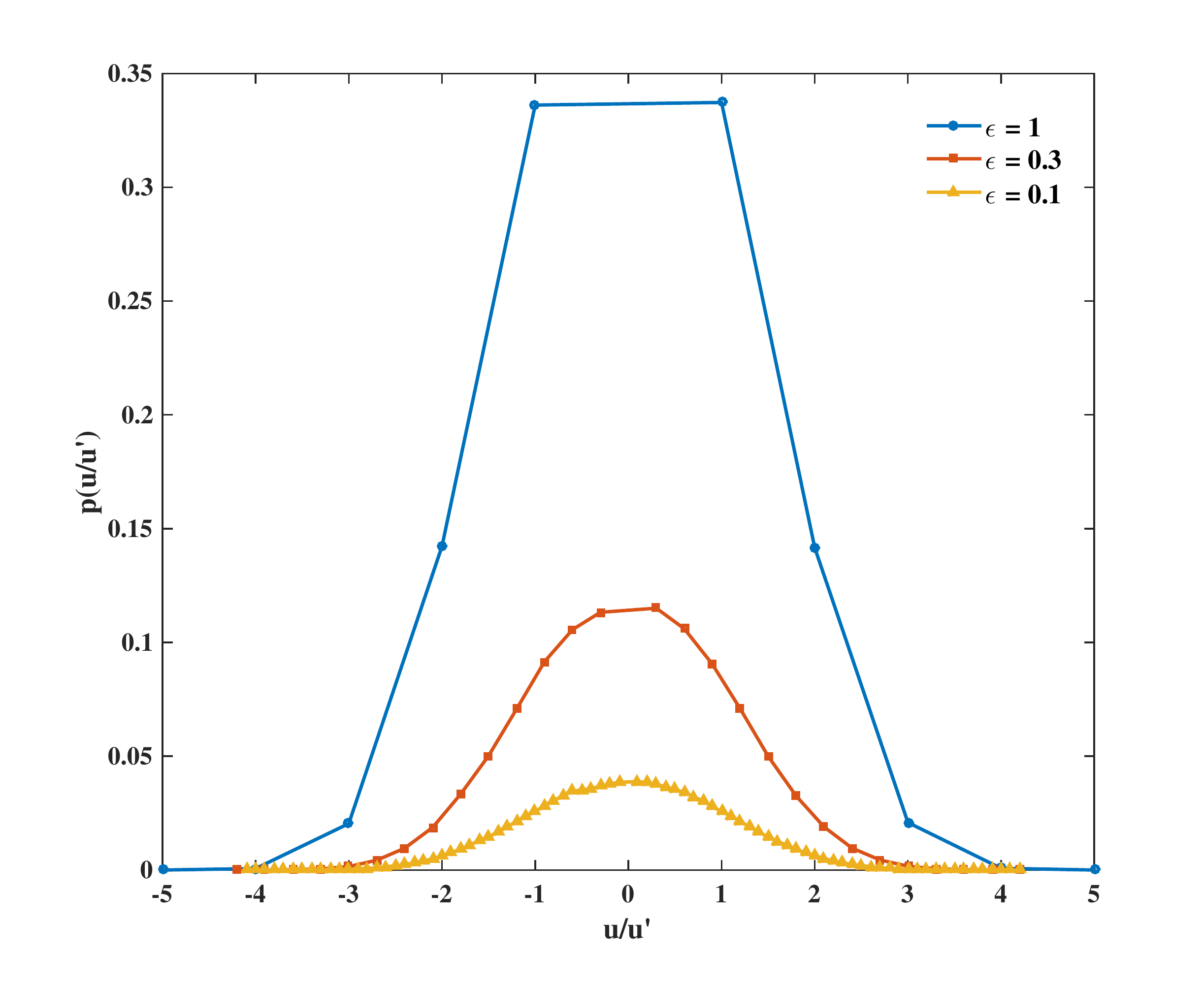}
\caption{Probability distributions (pdfs) of the velocity (after being normalized by the standard deviation $\sigma$, for three different bin sizes. Although the coarsest (largest binsize) pdf is less smooth, it still retains the general shape of the nearly gaussian distribution.}
\label{pdfs}
\end{figure}

Figure \ref{partition} shows how this binning works. Calling $\epsilon$ the bin size, a  continuous stream of data is converted to discrete data or symbols, each representing a range of values. If $\epsilon$ is small, then the number of symbols, the alphabet size, $A(\epsilon)$ is large and vice-versa. To be examined is the effect of varying $\epsilon$ on the pdf, and hence $H$. Figure \ref{pdfs} shows $u$ for several values of $\epsilon$. The general shape is the same, but the gaussian character is not clear but for small $\epsilon$.

What value of $\epsilon$ should be used? This is not an easy question to answer. It is tempting to try the limit $\epsilon \rightarrow 0$, as is done when calculating $e.g.$ the Kolmogorov-Sinai entropy \cite{frig2004}, but it should be clear that this is not an option for real data. An additional complication is noise, as highlighted by the laminar portion of Fig. \ref{slug}. If $\epsilon$ is below the noise, one gets a different value from the theoretical $H(U) = 0$.

It turns out that it may not matter, depending on the question one is trying to answer. Arbitrarily, we choose  $\epsilon = \sigma$, the standard deviation for the laminar noise and the turbulent slugs respectively (separately) and proceed to see how information theory can distinguish between the two. This corresponds to the coarsest pdf in Fig. \ref{pdfs}. If we use a different $\epsilon$, the results are qualitatively the same. The pdfs of both of the noisy laminar data and the turbulent data are nearly gaussian and will nearly collapse when normalized $\sigma$.  As a result, the two values of $H$ are nearly identical.

\begin{figure}[h!]
\hspace{-1.5em}
\centering
\includegraphics[scale = 0.4]{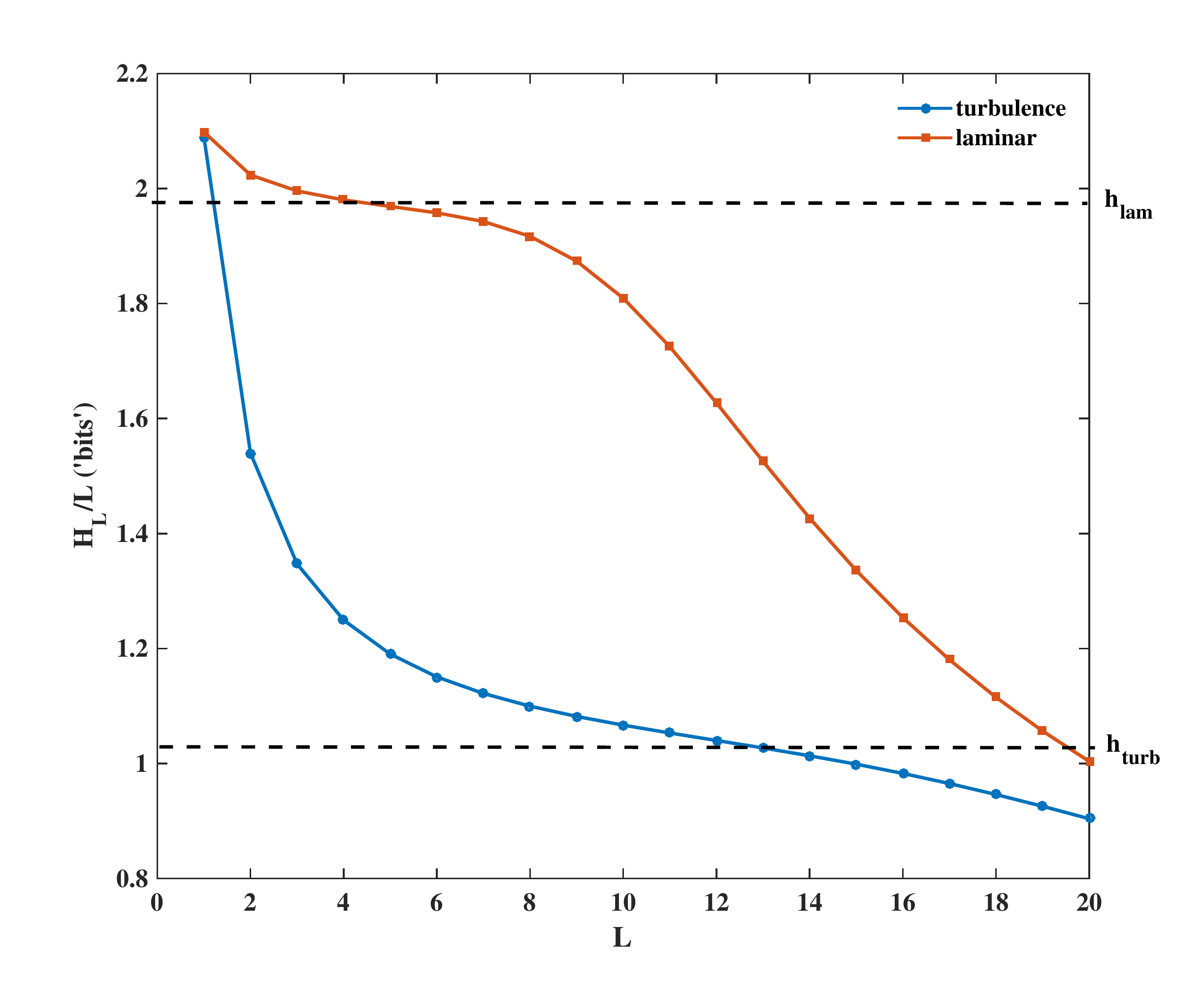}
\caption{The entropy densities $h_L  = H_L/L$ for laminar noise and turbulent slugs at $Re \approx 7000$. As $L$ increases, the correlations in turbulence quickly allow this measure to distinguish it from the shot noise in the laminar time series. $h_L$ will continue to decrease until all correlations are accounted for, or until the finite data length decreases it to zero artificially. The switch between these two effects, the inflection point, is typically considered the as the entropy density $h$. Here the two inflection points $h_{lam}$ and $h_{turb}$ are vastly different, speaking to the different character of the system.}
\label{slug_h}
\end{figure}   

True noise (often called shot noise \cite{vanetten2006}) is uncorrelated with itself at all finite lag times. An even stronger statement is to say that it is statistically independent of itself at different times. We can take advantage of this fact by considering finite blocks of the velocity time series: $u^L = [u(t),u(t+\Delta t),...,u(t+(L-1)\Delta t)]$, where $\Delta t$ is the inverse mean sampling rate of the LDV. The probabilities of these blocks directly relate to the inter-relationships between its members.

Consider the so-called  block entropy \cite{cover1991}:
\begin{equation}
H_L = H_L(U) = H(U^L)= -\sum_{u^L} p(u^L) \log p(u^L).
\label{Hblock}
\end{equation}
If there is no statistical dependence between any of the data points inside the blocks (no correlations) then $H_L(U) = H(U^L) = L H(U)$. By an application of Jensen's inequality \cite{cover1991}, this is the maximum value: $H_L(U) \le L H(U)$. It is lowered by any statistical dependence between the $U$'s inside of $U^L$, $i.e.$ temporal or spatial correlations. This observation can be exploited to distinguish the shot noise, seen even in laminar flows.

The above definition is applied to  $H_L$  for laminar and turbulent data from the pipe measurements. It is divided by $L$, so that if the data is truly random, it will return to the value $H(U^1)$. As expected, the random noise in the laminar flow has no correlations, so it is hardly reduced at all by this division, while the correlations inherent in the Kolmogorov picture of turbulence \cite{frisch1995} reduce the entropy for the turbulent flow, as shown in Fig. \ref{slug_h}.

It is instructive to examine how $H_L$ depends on $L$. This function is not well-behaved at very large $L$.  Let $L_0$ be the total number of velocity data points in a typical run (Typically, $L_0$ is of order $10^6$.) When the number $L_0/A(\epsilon)^L$ is too small, there are too few blocks to determine  the occurrence probabilities needed to evaluate $H_L$, resulting in the data appearing to be {\it less} random than it really is. This effect is compounded if velocities are correlated, as in turbulence, where the correlation length can be large and a large $L$ is needed to capture the physics.

Figure \ref{slug_h}, shows $H_L/L$ $vs$ $L$ for the turbulent fluctuations in slugs and the shot noise in the laminar flow. The laminar measurements show a dependence on $L$ for the physically uninteresting reason described above that the data are not collected for a long enough time interval to accurately measure probabilities appearing in $H_L$. This same phenomenon plagues the slug curve, and so researchers typically identify the inflection point as being important \cite{cerbus2013}. If $H_L/L$ did not drop off, it would remain constant and this limit is called the entropy rate or density \cite{cover1991}:
\begin{equation}
h = \lim_{L \rightarrow \infty} h_L \equiv \lim_{L \rightarrow \infty} \frac{H_L}{L}.
\end{equation}
There are three equivalent ways to define $h$. The first is described above, and the second is
\begin{equation}h =  \lim_{L \rightarrow \infty}(H_{L+1}  - H_L ),
\end{equation}
while the third is
\begin{equation}
h =  \lim_{L \rightarrow \infty}H( X^1 | X^L).
\label{entropyDensity}
\end{equation}
Here $H(X|Y)$ is the conditional entropy ($X$ conditioned on $Y$). The entropy density converges to a finite value because less and less information is gained on increasing $L$ by one unit. The utility of $h$ in elucidating turbulence will be explored further in the soap film experiments to be described next.

\section{Soap film: Quasi-2D turbulence}

\begin{figure}[h!]
\hspace{-1.5em}
\centering
%\centerequivalently
\includegraphics[scale = 0.23]{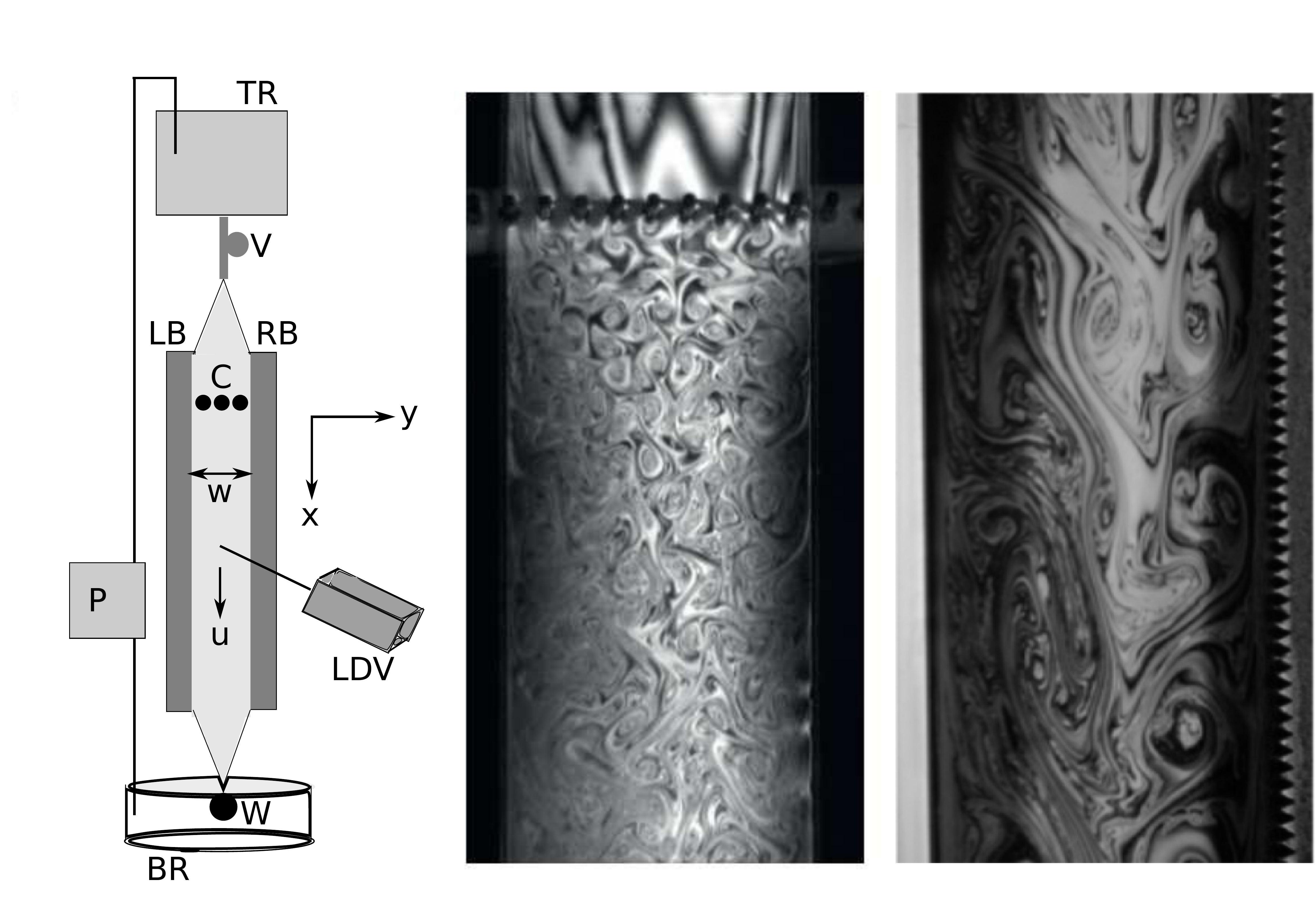}
\caption{Left: Experimental setup showing the reservoirs ($TR$, $BR$), pump ($P$), valve ($V$), comb ($C$), and blades ($LB$, $RB$). Center: An interference pattern created by shining monochromatic light on the turbulent soap film. The fluctuations in film thickness which give rise to the thin film interference are driven by the turbulent velocity field. The turbulence is generated by inserting a comb or using a rough wall. The comb initiates vortices that break up into smaller ones, a phenomenon call the ``enstrophy cascade''.  The rough wall causes the opposite effect; it initiates a cascade of energy to larger and larger sizes, until they become comparable to the width of the film or even larger. This is called the inverse energy cascade case, a phenomenon that cannot occur in typical three-dimensional flows.}
\label{setup}
\end{figure}

The gravity-driven soap film is a mixture of soap detergent and water, with small density-matched glass spheres added for the velocity measurements. Fig. \ref{setup} is a diagram of the experimental setup. A Laser Doppler velocimeter (LDV) is used to measure the velocity components in the streamwise $u$ direction. By adjusting the flow rate and the channel width, several decades of Reynolds number $Re$ can be explored. Here $Re = u'w/\nu$, where $u'$ is the rms velocity and $w$ is the channel width. For more experimental details, we refer the reader to Ref. \cite{cerbus2013}.

\begin{figure}[h!]
\hspace{-1.5em}
\centering
\includegraphics[scale = 0.4]{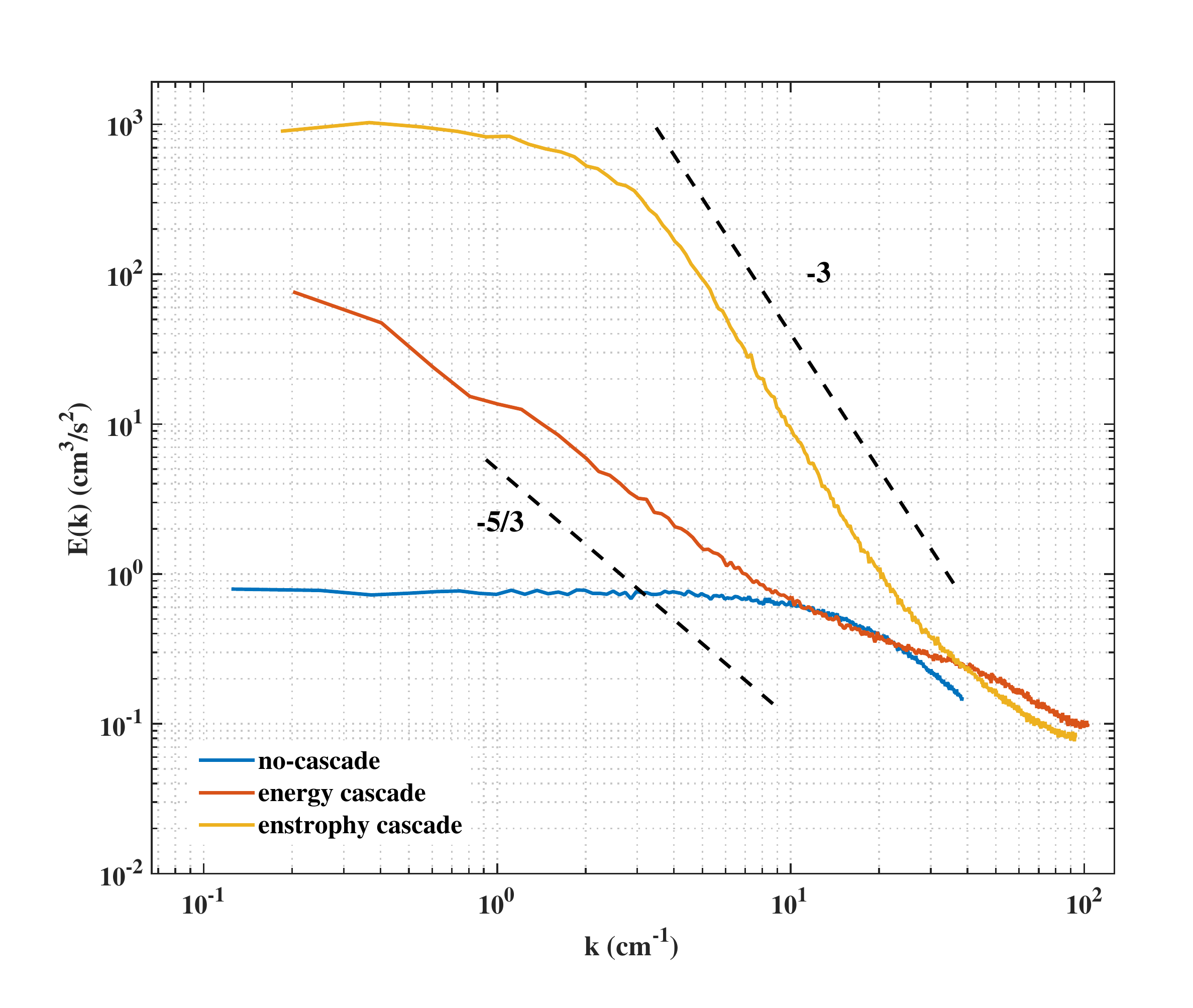}
\caption{The canonical spectra of 2D turbulence representative of the three regions shown in Fig. \ref{h_L}. For the enstrophy cascade, $E(k) \propto k^{-3}$, and for the inverse energy cascade, $E(k) \propto k^{-5/3}$ \cite{kellay2002,boffetta2012}. For low $Re$ $E(k)$ is flat, indicating that no cascade is present, but the flow is not laminar. The tail of all three curves are strongly affected by the random, finite sampling rate of the LDV \cite{adrian1986}.}
\label{spectra} 
\end{figure}

Using Eq. \ref{entropyDensity}, we go on to calculate the entropy rate for turbulence in a soap film. Here, depending on the forcing, three entirely different kinds of behavior can be generated. If the soap film travels between rough walls, the perturbation they steadily create an inverse cascade of energy from smaller to larger eddies \cite{kellay2002,boffetta2012}. If, instead, the film is penetrated by a comb, a row of $\simeq$ 1mm rods, a different type of cascade is created; the rods create vortices which cascade downscale to the smallest sustainable size: the direct enstrophy cascade \cite{kellay2002,boffetta2012}. The accompanying energy spectra $E(k)$ are as seen in Fig. \ref{spectra}. Here $k$ is wavenumber in inverse cm and $E(k)$ has units of kinetic energy per kg per unit wave number. A final case is when the perturbations of the comb are too weak to initiate a cascade at all, resulting in a flat $E(k)$.

\begin{figure}[h!]
\hspace{-1.5em}
\centering
\includegraphics[scale = 0.4]{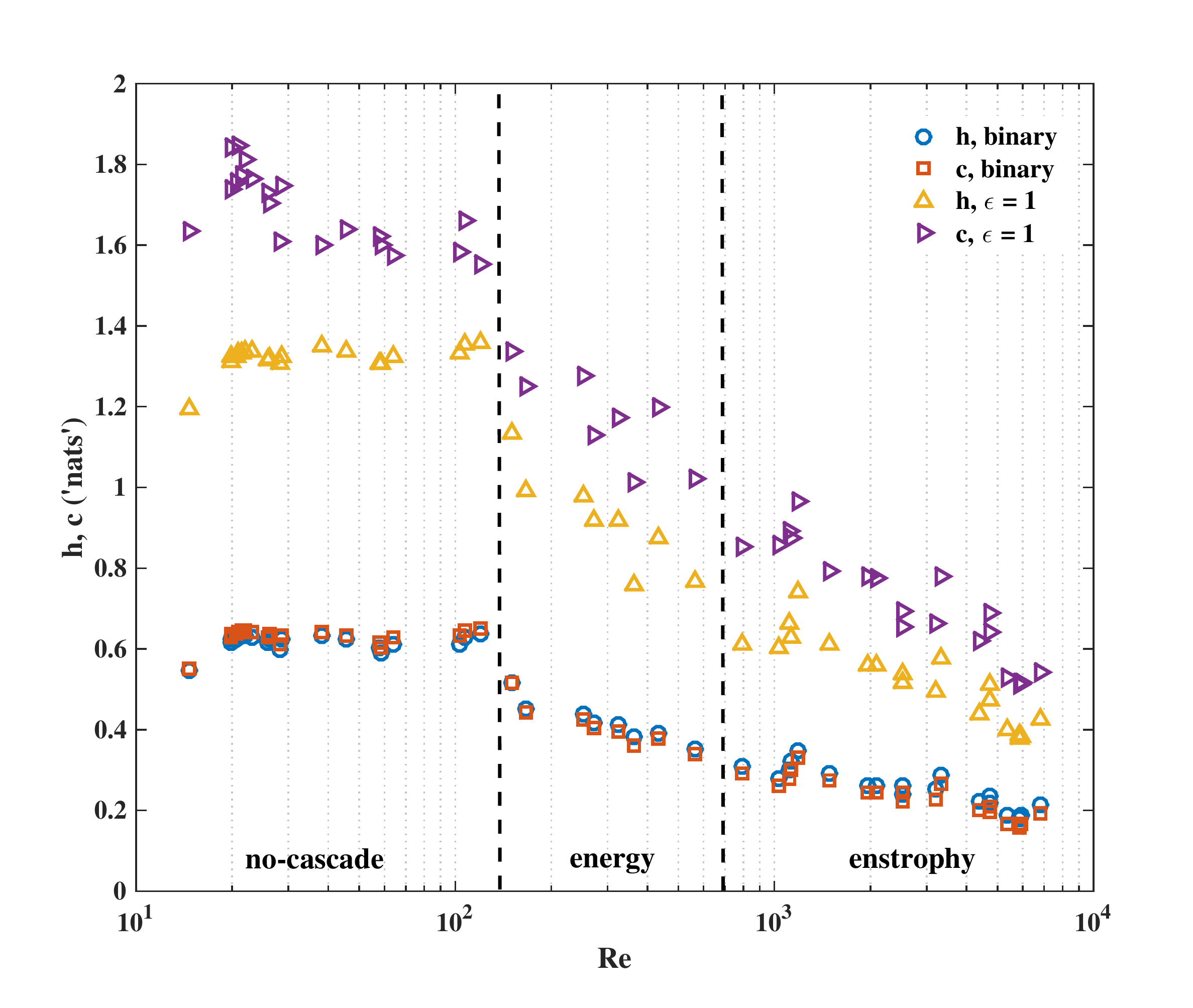}
\caption{The entropy density $h$ calculated from experiments carried out in a turbulent soap film plotted along with $c$ the information content deduced using the data compression recipe of Ziv and Lempel. All velocities were normalized by the standard deviation $\sigma$ before processing to achieve nearly equal alphabet sizes $A(\epsilon)$. In one set of measurements the data are binarized (open squares and open circles). It is not surprising that this crude treatment of the data shows a smaller value of the entropy density $h$, since since less information about the data is conveyed than if they are segmented on a decimal scale to one significant figure (0,0.1, 0.2., ..., 0.9).}
\label{h_L} 
\end{figure}

Besides $h$, another quantity is also plotted in Fig. \ref{h_L}. This is an alternative method for estimating the information and in the limit of infinite data the two coincide. Computer memory is necessarily limited, making it useful to store and transmit information in as compact a form as possible. The total amount of memory necessary to send or store a message can be shortened by (re-)coding the words to minimize their length. One example is using a coding scheme that assigns a small number symbols to words that appear with high frequency, as in Morse code where $e$ is coded with the shortest symbol, a 'dot'. In turbulence the probability distribution like in Fig. \ref{pdfs} suggests something similar might be done. Remarkably, Shannon proved that the limit on $any$ compression scheme is given by the entropy of the message \cite{cover1991}. If the original length or size of a message is $S_0$, then after the compression algorithm does its work, the new size $S$ satisfies the inequality
\begin{equation}
\frac{S}{S_0} \le h.
\end{equation}
While Shannon provided no hints as to how equality might be approached, substantial work has been done in his wake to satisfy the equality in the limit $S_0 \rightarrow \infty$. These coding schemes are called optimal, an example being the Lempel-Ziv algorithm \cite{schurmann1996,ziv1978}. As expected, the compression ratio $c$ in Fig. \ref{h_L} is equal to or greater than $h$; it should not be smaller. Binarizing the data gives a rough description, so it is not surprising that the difference between $c$ and $h$ is revealed only when the data is segmented into ten values. 

Why does $h$ decrease as $Re$ increases? Presumably this because increasing the strength of the turbulence increases the correlations between different spatial scales, and correlations always $decrease$ the surprise element in all observations \cite{cover1991}. But a puzzle remains: when $Re$ is decreased sufficiently, the flow must become laminar, in which case $h$=0. Therefore $h$ must go through a maximum - at a value that these experiments cannot measure.         

A strange prediction is suggested by Fig. \ref{h_L}. If we treat $h(Re)$ as a state function in thermodynamics, then changing $Re$ simply moves the system along this unique curve. Consider the case of decaying turbulence, as with a comb in a soap film. Near the comb the flow exhibits the enstrophy cascade and $Re$ as defined here is large. Measurements made progressively further away from the comb have a lower energy due to the decay, and so a lower $Re$. At the same time, $h$ will increase until there is a transition, apparently continuous with respect to $h$, from an enstrophy cascade to an inverse energy cascade. This remarkable transition has, in fact, been observed in soap films, although the mechanism has been attributed to wall shear effects \cite{liu2016}.

\section{GOY Model of Turbulence}

To further illustrate some of the tools of information theory in practice, we turn to a toy model of turbulence designed to mimic the essential properties of turbulence. The Gledzer-Ohkitani-Yamada (GOY) shell model is the simplest model with a cascade of energy \cite{pisarenko1993,kadanoff1995,kadanoff1997}, but it still yields to theoretical analysis and can easily be numerically integrated even on a laptop computer. Herein lies its utility. As long as its limits are kept in mind, the GOY model can be a useful playground for new ideas that can later be applied to full-blown Navier-Stokes turbulence. This approach has led to the discovery of intermittency in the helicity cascade of 3D turbulence \cite{chen2003,kadanoff1995}. As Kadanoff quipped, "Models are fun, and sometimes even instructive" \cite{kadanoff1995}.

In the GOY model, each variable (shell) corresponds to velocity fluctuations on a different spatial scale. These shells, however, live in Fourier space, and so the independent variable is the wavenumber $k$. (This is also why the velocity of the shells are complex). It is useful to think of this model as a truncation of the Navier-Stokes equation in Fourier space. 

There are a finite number ($N$) of shells and we denote by $n$ any particular shell (from 1 to $N$). Following custom, $N$ is set to 22 \cite{materassi2014,kadanoff1995b}. The wavenumbers $k_n$ are picked to be a fixed logarithmic distance apart: $k_n = k_0 q^n$, where here $k_0 = 2^{-4}$ and $q = 2$. Large $n$ corresponds to small scales, while small $n$ refers to large scales. The governing set of equations is
\begin{equation} 
{\Big (} \frac{d}{dt} + \nu k {\Big )} u_n(t) = i(a_n u_{n+1}(t) u_{n+2}(t) + b_n u_{n-1}(t) u_{n+1}(t) + c_n u_{n-1}(t) u_{n-2}(t))^* + f_n
\end{equation}
where $f_n$ is the forcing, $i = \sqrt{-1}$, and $^*$ denotes the complex conjugate. The variables $a_n$, $b_n$ and $c_n$ are shell dependent but constant in time. These determine the strength of energy flow between scales. We refer the reader to the accessible review by Kadanoff \cite{kadanoff1995} or the detailed review by Biferale \cite{biferale2003} for more particulars. Following the numerical scheme outlined in Ref. \cite{pisarenko1993}, the  viscosities are assigned the values $\nu = 10^{-7}$, $10^{-6}$, $10^{-5}$, where the Reynolds number $Re \propto 1/\nu$.  The equations are integrated using MATLAB. The energy spectra are shown below in Fig. \ref{goy_spectra}. As the viscosity decreases ($Re$ increases), the Kolmogorov -5/3 scaling is present at higher and higher wavenumbers.

\begin{figure}[h!]
\centering
\includegraphics[scale = 0.4]{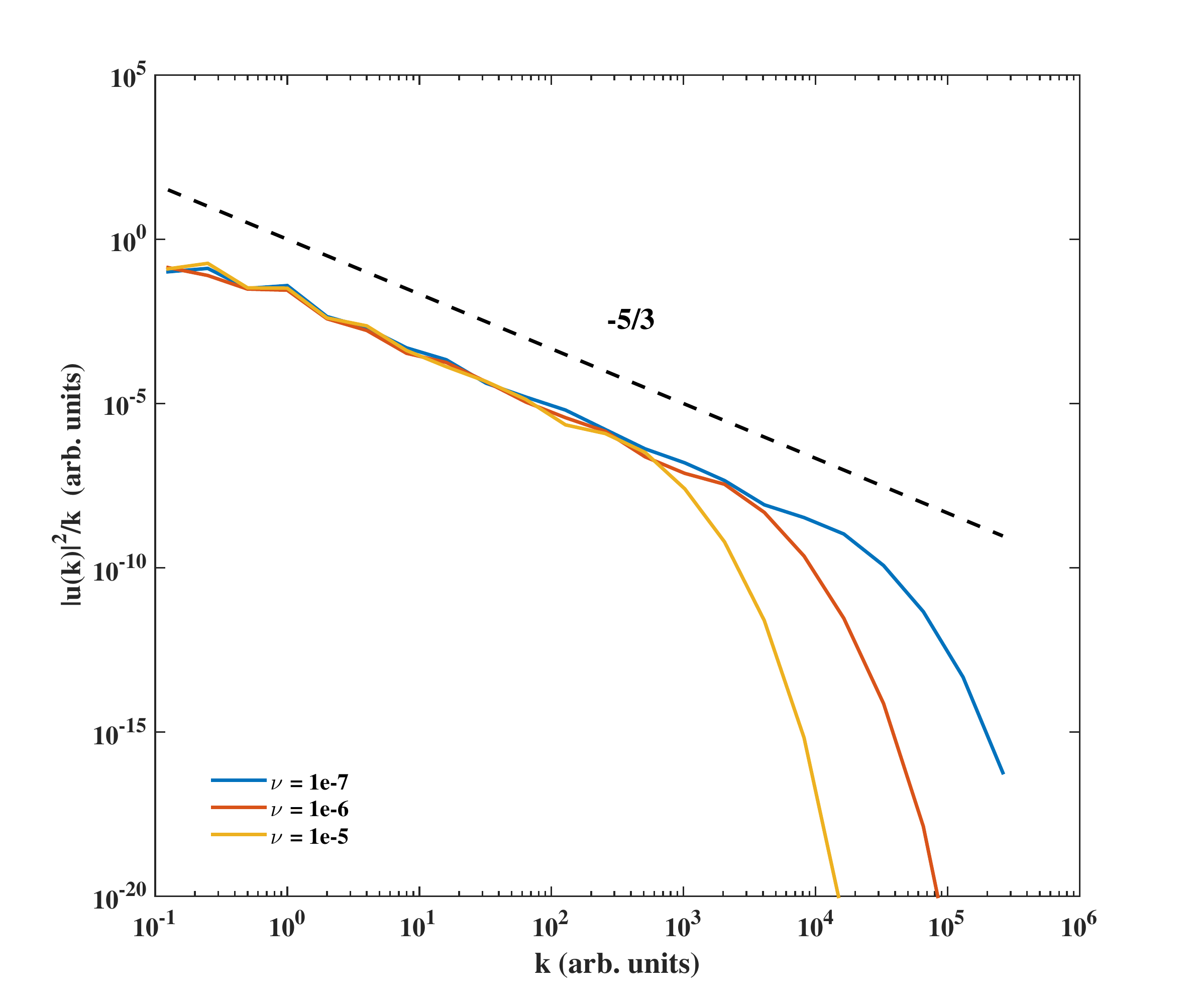}
\caption{The energy spectra for the GOY model for viscosities of $\nu = 10^{-7}$, $10^{-6}$, $10^{-5}$. The spectra show a wide (-5/3) inertial range by construction, with a dissipative cutoff at higher $k$ for higher $1/\nu$.}
\label{goy_spectra} 
\end{figure}   

\begin{figure}[h!]
\centering
\includegraphics[scale = 0.4]{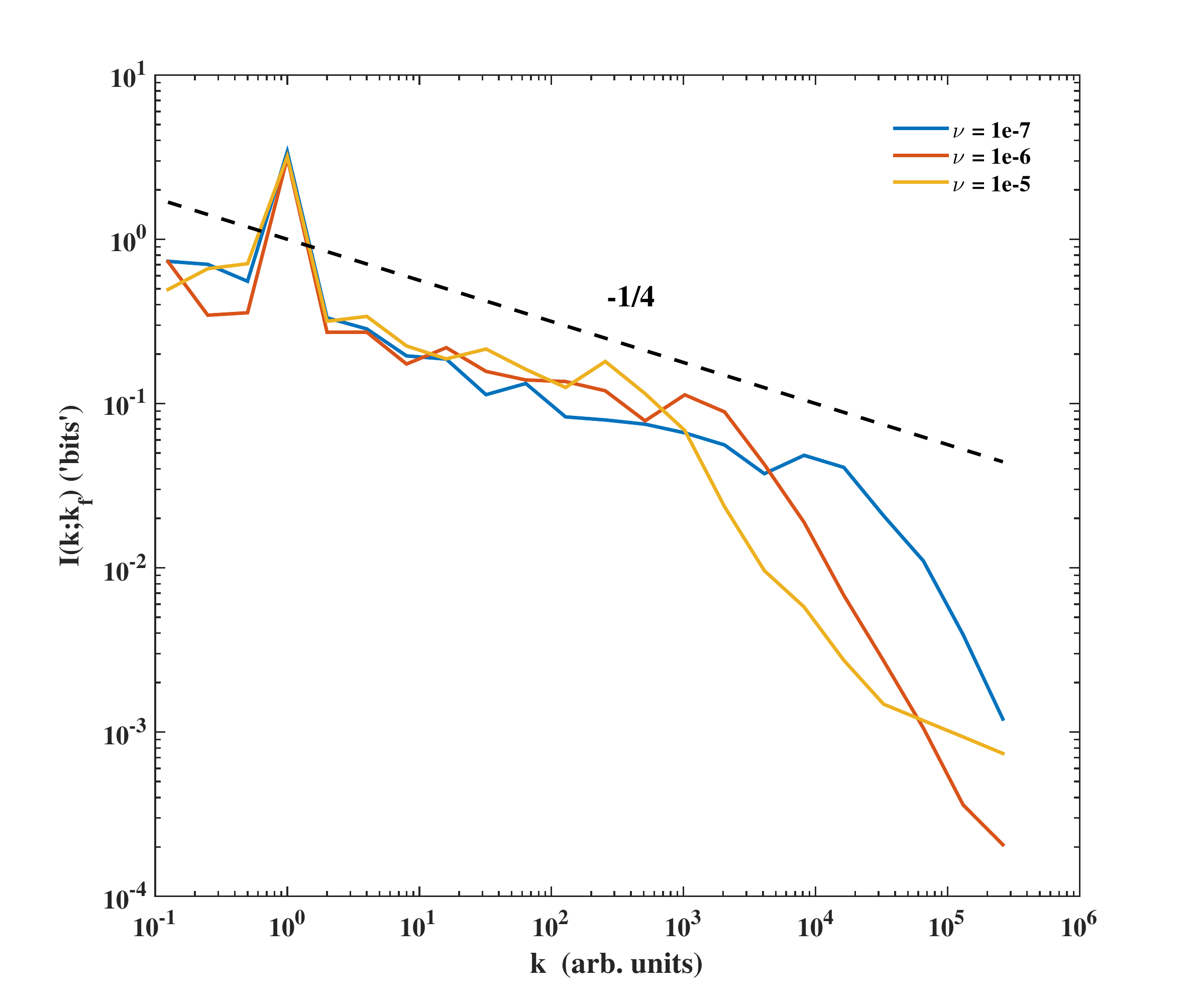}
\caption{The mutual information between the forcing scale ($n = 4$) and all other scales for the same viscosities as in Fig. \ref{goy_spectra}. The -1/4 power is not understood, but it implies that there is no characteristic distance from the forcing at which the input is forgotten.}
\label{forcing_mutinfo} 
\end{figure}

What sort of questions can be address  with the tools of information theory?  Consider one of the most fundamental concepts in turbulence phenomenology, the universality of small scales. In the traditional picture of the turbulent cascade, energy is injected at some large scale and then transferred to smaller and smaller scales. In order for the scaling of the velocity differences to depend only on the local scale $1/k$ and the energy injection rate, it must ``forget'' about the large scales. Or, as Kadanoff puts it, the system remembers as little as it possibly can about the conditions at which energy is added or dissipated \cite{kadanoff1997}.  This forgetfulness is readily quantified using information theory, in particular through the mutual information. If one scale forgets about the other, then the mutual information between them will be small (identically zero if statistically independent):
\begin{equation}
I(e(k); e(k + \Delta k)) = I(k; k + \Delta k),
\end{equation}
where $e(k) = |u(k)|^2$ is the energy at $k$ and $\Delta k$ is the spacing between adjacent shells. Figure \ref{forcing_mutinfo} is a plot the instantaneous mutual information between the energy each scale and at the forcing scale ($n$ = 4 shell). Apart from the peak at $n$ = 4, where the mutual information is now equal to the self-information (the entropy), the shared information falls off algebraically as -1/4 until the dissipative scales are reached. Thus the inertial scales do not completely forget the details of the forcing, although the shared information becomes very small. Indeed, the GOY model is well known to exhibit the same intermittency phenomenon (deviations from Kolmogorov and violations of the universality assumption) as fluid turbulence \cite{kadanoff1995b}.

While the entropy plays the most fundamental role, that of a measure, in Shannon's information theory, it is the mutual information that has proved the most useful in the field of communications. In any causal physical relationship, or interaction, one system or subsystem will exchange with another not only energy or momentum but also information. (Indeed, it is well known that the speed of light restriction of Special Relativity is most properly formulated in terms of information transfer, albeit not explicitly in the Shannon sense.)

It is reasonable to consider whether the cascade may also be a communication channel. To do so, let us now consider the local transfer of information between adjacent shells using the mutual information. Since the information source, just like the energy, must be at the forcing scale, we expect information to flow downscale. However, it is well known that there is ``backscatter'' of energy (enstrophy) even in the 3D energy (2D enstrophy) cascade. With this in mind, we also consider the transfer of information upscale, as done in \cite{materassi2014}. 

\begin{figure}[h!]
\centering
\includegraphics[scale = 0.4]{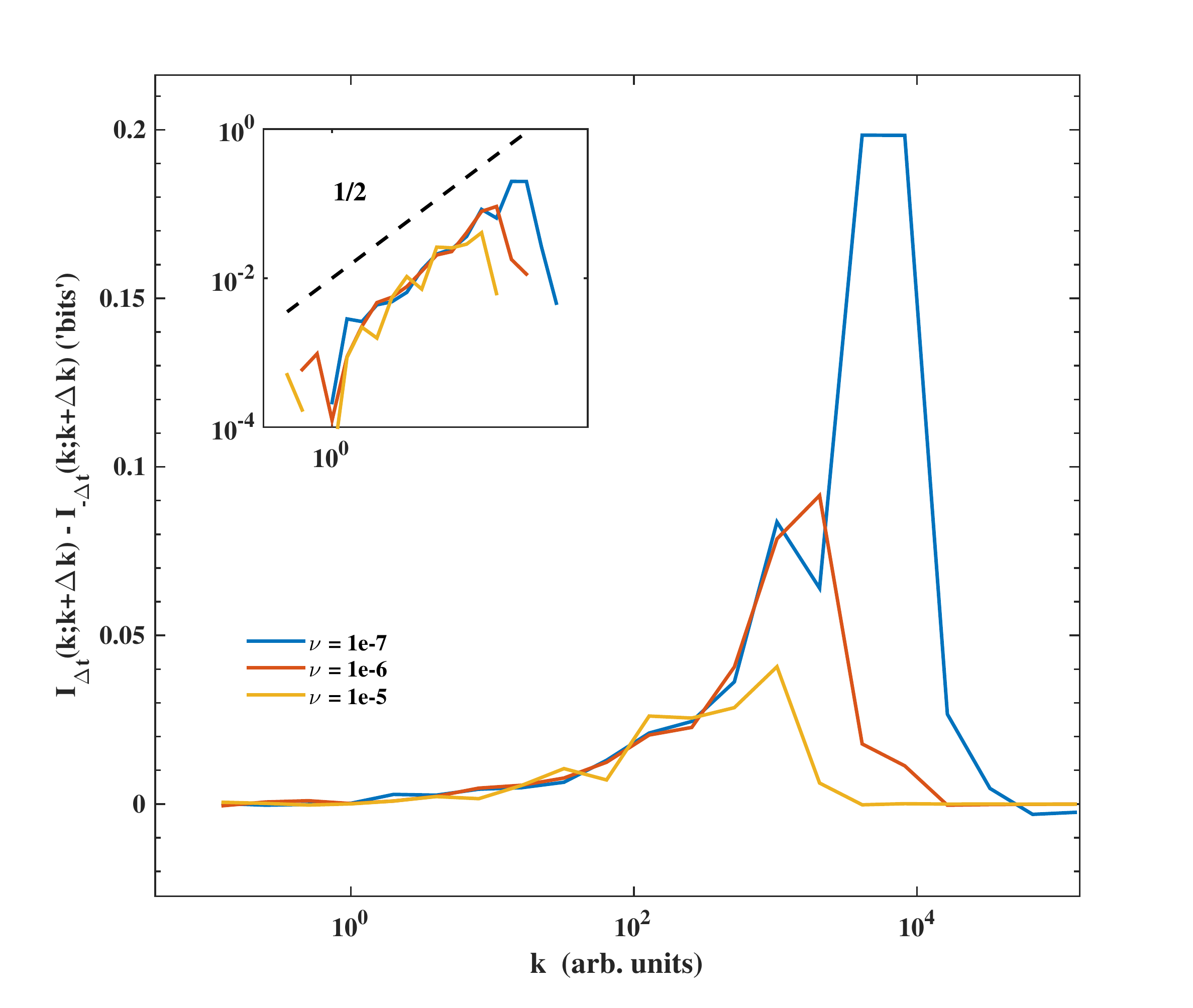}
\caption{The mutual information between adjacent scales with a time lag $\Delta t$ arranged so that a positive difference means a net transfer of information downscale. The viscosities are the same as in Fig. \ref{goy_spectra} and $\Delta t$ was arbitrarily chosen as 100 integration time steps, equivalent to $\approx 0.002$ turnover times \cite{pisarenko1993}. Concurrent with the cascade of energy is an information cascade that grows with an unexplained power law of $1/2$ in the inertial range before being cutoff by viscosity.}
\label{info_transfer} 
\end{figure}

The cascade is a dynamical process. An amount of energy at a wavenumber $k$ at time $t$ is transferred to wavenumber $k + \Delta k$ in a time $\Delta t$. To reflect this, we introduce a time lag into the mutual information. A large shell (``eddy") at an arbitrary time $t$ will share information forward in time to a small shell, so
\begin{equation}
I(e(k,t); e(k + \Delta k,t + \Delta t))
\end{equation}
will be large if information is going downscale, whereas the following will be large if information is going upscale
\begin{equation}
I(e(k+\Delta k,t);e(k,t+\Delta t)) = I(e(k,t+\Delta t);e(k+\Delta k,t))
\end{equation}
The use of this time lag is similar to its use in \cite{vastano1988,schreiber2000,materassi2014}. Subtracting the transfer upscale from that downscale, in Fig. \ref{info_transfer} we indeed find a net downscale transfer that is positive only in the inertial range and increases in magnitude as $Re$ increases. Thus there exists a companion information cascade along with the energy cascade in the GOY model. It would be interesting to determine if the same is true for 3D and 2D turbulence. The presence of an information flux is not only related to intermittency, but suggests a further useful constraint on the physics of turbulence.

\section{Summary}

While information theory was created to characterize  messages coded in the form of words, it applies equally well to experimental observations in which the words of a text are replaced by measured quantities.   In this work, information is applied to three-dimensional turbulent flow in a long pipe, to  two-dimensional turbulence in a gravity-driven soap film, and to a mathematical model that takes into account the turbulent cascade of energy from larger to smaller eddy sizes.  The goal of the work is to demonstrate that information theory can illuminate physical observations,  even when the equations governing the system's behavior are  intractable or may not even be known.  In this study, no appeal is made to the Navier-Stokes equations, which govern the fluid flows under observation.  Even when true velocity fluctuations are absent, as in laminar flows, shot noise (also called Poisson noise), can appear as a confounding effect.  The goal of the work is to introduce the reader to the information theory approach, and to demonstrate its usefulness.  

\begin{acknowledgements}
This work is supported by NSF Grant No. 1044105 and by the Okinawa Institute of Science and Technology (OIST).
\end{acknowledgements}

\end{document}